\documentclass{emulateapj}

\slugcomment{Submitted 13$^{\rm th}$ March 2008}

\shorttitle{Mid-Infrared Spectra of High Redshift ($z>2$) Radio Galaxies}
\shortauthors{N. Seymour et al.}

\def\arcsec{\ifmmode {^{\prime\prime}}\else $^{\prime\prime}$\fi}
\def\arcmin{\ifmmode {^{\prime}}\else $^{\prime}$\fi}
\def\farcs{\ifmmode \rlap.{^{\prime\prime}}\else
    $\rlap.{^{\prime\prime}}$\fi}
\def\arcmper{\ifmmode \rlap.{^{\prime}}\else
    $\rlap.{^{\prime}}$\fi}

\def\lya{\ifmmode {\rm Ly\alpha}\else{\rm Ly$\alpha$}\fi}
\def\Lya{\ifmmode {\rm Ly\alpha}\else{\rm Ly$\alpha$}\fi}
\def\LFIR{\ifmmode {\rm \,L_{FIR}}\else ${\rm \,L_{FIR}}$\fi}
\def\Lsun{\ifmmode {\rm\,L_\odot}\else ${\rm\,L_\odot}$\fi}
\def\Msun{\ifmmode {\rm\,M_\odot} \else ${\rm\,M_\odot}$\fi}
\def\Msunpyr{\ifmmode {\rm\,M_\odot\,yr^{-1}} \else {${\rm\,M_\odot\,yr^{-1}}$}\fi}
\def\pyr{\ifmmode {\rm\,yr^{-1}} \else {${\rm\,yr^{-1}}$}\fi}
\def\kms{\ifmmode {\rm\,km~s^{-1}} \else ${\rm\,km\,s^{-1}}$\fi}
\def\kmps{\ifmmode {\rm\,km~s^{-1}} \else ${\rm\,km\,s^{-1}}$\fi}
 
\def\ergps{\ifmmode {\rm\,erg\,s^{-1}} \else {${\rm\,erg\,s^{-1}}$}\fi}
\def\ergpspcm{\ifmmode {\rm\,erg\,s^{-1}\,cm^{-2}} \else {${\rm\,erg\,s^{-1}\,cm^{-2}}$}\fi}
\def\surfbr{\ifmmode {\rm\,erg\,s^{-1}\,cm^{-2}\,arcsec^{-2}} \else {${\rm\,erg\,s^{-1}\,cm^{-2}\,arcsec^{-2}}$}\fi}
\def\spose#1{\hbox to 0pt{#1\hss}}
\def\simlt{\mathrel{\spose{\lower 3pt\hbox{$\mathchar"218$}}
     \raise 2.0pt\hbox{$\mathchar"13C$}}}
\def\simgt{\mathrel{\spose{\lower 3pt\hbox{$\mathchar"218$}}
     \raise 2.0pt\hbox{$\mathchar"13E$}}}

\def\eg{{\it e.g.,}}

\def\erg{{\rm\thinspace erg}}

\def\uJy{{\rm\thinspace \mu Jy}}

\def\km{{\rm\thinspace km}}

\def\Mpc{{\rm\thinspace Mpc}}

\def\s{{\rm\thinspace s}}

\def\ergps{\mbox{$\erg\s^{-1}$}}

\def\kmps{\hbox{$\km\s^{-1}\,$}}

\def\kmpspMpc{\hbox{$\km\s^{-1}\Mpc^{-1}$}}

\def\um{\hbox{$\mu {\rm m}$}}

\def\Msun{{\rm\thinspace M_\odot}}

\begin{document}

\title{Mid-Infrared Spectra of High Redshift ($z>2$) Radio Galaxies}

\author{N. Seymour\altaffilmark{1}, 
  P. Ogle\altaffilmark{1},
  C. De Breuck\altaffilmark{2},
  G. G. Fazio\altaffilmark{3},
  A. Galametz\altaffilmark{2,4}
  M. Haas\altaffilmark{5},
  M. Lacy\altaffilmark{1},
  A. Sajina\altaffilmark{6},
  D. Stern\altaffilmark{4},
  S. P. Willner\altaffilmark{3},
  J. Vernet\altaffilmark{2},
}

\altaffiltext{1}{{\it Spitzer} Science Center, California Institute of
  Technology, 1200 East California Boulevard, Pasadena, CA 91125, USA. 
[email: {\tt seymour@ipac.caltech.edu}]}
\altaffiltext{2}{European Southern Observatory, Karl Schwarzschild Stra\ss e, 
D-85748 Garching, Germany.}
\altaffiltext{3}{Harvard-Smithsonian Center for Astrophysics, 60 Garden Street, Cambridge, MA 02138, USA.}
\altaffiltext{4}{Jet Propulsion Laboratory, California Institute of Technology, Pasadena, CA 91109, USA.}
\altaffiltext{5}{Astronomisches Institut, Ruhr-Universit\"at Bochum, Universit\"atsstr. 150, Geb\"aude NA 7/173, D-44780 Bochum, Germany.}
\altaffiltext{6}{Departments of Physics \& Astronomy, Marian E. Koshland Integrated Natural Science Center, 370 Lancaster Avenue, Haverford College, Haverford, PA 19041, USA.}

\begin{abstract}
We present the first mid-infrared {\it Spitzer}/Infrared Spectrograph (IRS)
observations of powerful radio galaxies at $z$$>$$2$. These radio 
galaxies, 4C\,+23.56 ($z$=$2.48$) and 6C\,J1908+7220 ($z$=$3.53$), both show strong 
mid-infrared continua, but with 6C\,J1908+7220 also showing strong PAH emission 
at rest-frame 6.2 and $7.7\,\um$. In 4C\,+23.56 we see no obvious PAH features 
above the continuum. The PAH emission in 6C\,J1908+7220
is the amongst the most distant observed to date and implies that there is a large 
instantaneous star formation rate (SFR). This is consistent with the
strong detection of 6C\,J1908+7220 at far-IR and sub-mm wavelengths, indicative
of large amounts of cold dust, $\sim$$10^9\,M_\odot$. Powerful radio galaxies 
at lower redshifts tend to have weak or undetectable PAH features 
and typically have lower far-IR luminosities. In addition, 4C\,+23.56 shows 
moderate silicate absorption as seen in less luminous radio galaxies, indicating 
$\tau_{9.7\,\um}$=$0.3\pm0.05$. This 
feature is shifted out of the observed wavelength range for 6C\,J1908+7220. 
The correlation of strong PAH features with large amounts of cold dust,
despite the presence of a powerful AGN, is in agreement with other recent 
results and implies that star formation at high redshift is, in some cases at 
least, associated with powerful, obscured AGN.
\end{abstract}

\keywords{galaxies: active ---  galaxies:  high-redshift --- galaxies:
evolution -- radio galaxies: individual(4C\,+23.56, 6C\,J1908+7220)}

\section{Introduction}

There is now clear evidence that high redshift, powerful radio galaxies 
(HzRGs; $z>1$ and $L_{\rm 3\,GHz} > 10^{26}\, {\rm W}\,$Hz$^{-1}$) have massive, 
$\sim 5 \times 10^{11}$\, M$_{\odot}$, 
stellar hosts (Seymour et al. 2007; see also Lilly
\& Longair 1984; De Breuck et al. 2001; Rocca-Volmerange et al. 2004)
\nocite{Lilly:84, DeBreuck:01, RoccaVolmerange:04, Seymour:07} and that 
the stellar mass of these hosts is related to the radio power of the HzRGs
\citep{Lacy:00, Willott:03}.
The strong radio and mid-IR {\em continuum} 
luminosities of these rare sources imply high accretion rates onto their 
central massive black holes  \citep{Ogle:06, Seymour:07}.  There is also 
evidence for extremely high SFRs from the strong 
sub-mm emission seen in some HzRGs \citep{Archibald:01, Reuland:04, Greve:06a},
as well as from the spectacular ($>100$\, kpc) luminous Ly$\alpha$ haloes 
observed in many HzRGs
\citep{Reuland:03, VillarMartin:03}. As the time scales of both the radio jets 
and star formation are very short ($\le0.1\,$Gyr), and as HzRGs are rare objects, 
it seems we are observing these sources at a special phase 
during a strong spurt of coeval black hole and galaxy growth, but when both 
processes are dust-enshrouded.

Aside from sub-mm/far-IR observations, the only other method to obtain an approximate 
measure of the SFR in HzRGs is from the mid-IR PAH emission features. 
This diagnostic is known to be inexact, especially in AGN, but UV/optical SFR diagnostics 
(e.g. narrow lines) are particularly difficult for HzRGs, as are radio/X-ray 
diagnostics. All these observables are known to be 
significantly contaminated by the powerful AGN. A correlation between PAH and sub-mm
luminosity has been found in local starbursts \citep{Haas:02, Peeters:04} showing that PAH 
emission can be used as an approximate tracer of star formation. 
There are now several results in the literature linking PAH emission and the presence
of cold dust associated with star formation (Lacy et al. 2007; Pope et al. 
2007; Sajina et al. 2008).\nocite{Lacy:07,Pope:07}

This letter presents {\it Spitzer}/IRS spectra of two $z>2$ HzRGs from our 
sample of 69 HzRGs studied by \citet{Seymour:07}: 4C\,+23.56 at $z$=$2.48$ and 
6C\,J1908+7220 (less commonly known as 4C\,+72.26) at $z$=$3.53$. 
Very few HzRGs have photometry at $24\,\um$ and even fewer are bright enough 
to be easily targeted by IRS. We select two of three HzRGs with 
$S_{\rm 24\um}>1\,$mJy 
which are also detected at $70\,\um$. While these HzRGs may not be 
representative of the entire population, even this small sample shows 
diverse mid-IR properties.

We also fit the total IR, $1-1000\,\um$, SEDs of these HzRGs 
and compare the model fits to the 
IRS spectra. We adopt the concordance $\Lambda$CDM 
cosmology with $\Omega_M$=$0.3$, $\Omega_\Lambda$=$0.7$, and 
$H_0$=$70\,\kmpspMpc$.

\section{The {\it Spitzer} Mid-IR Spectra}

The IRS spectra of 4C\,+23.56 were obtained as part of our Cycle 4 
GTO program (PI G. Fazio) and consisted of 12 ramp cycles of 120\,s each with 
just the LL1 module ($\sim20-36\,\um$). 6C\,J1908+7220 was observed as part of a 
Cycle 1 GO program (PI M. Haas)
and the data include 3\,cycles $\times60\s$ with the SL1 module, and 
3\,cycles $\times120\s$ with both the LL1 and LL2 modules.
We used the interactive {\tt IRSCLEAN}$^{\rm 1}$ task to remove bad pixels in 
the {\it Spitzer} post-BCD images and extracted the spectra with 
{\tt SPICE}\footnote{These codes are publicly available from the SSC website: 
{\tt http://ssc.spitzer.caltech.edu/postbcd/irs\_reduction.html}.} 
using `optimal' extraction. The spectra were smoothed by different amounts 
in each module varying from $1\,\um$ at the shortest wavelengths to 
$0.3\,\um$ at the longest wavelengths.

\begin{figure*}
\center
\includegraphics[width=7cm,angle=-90]{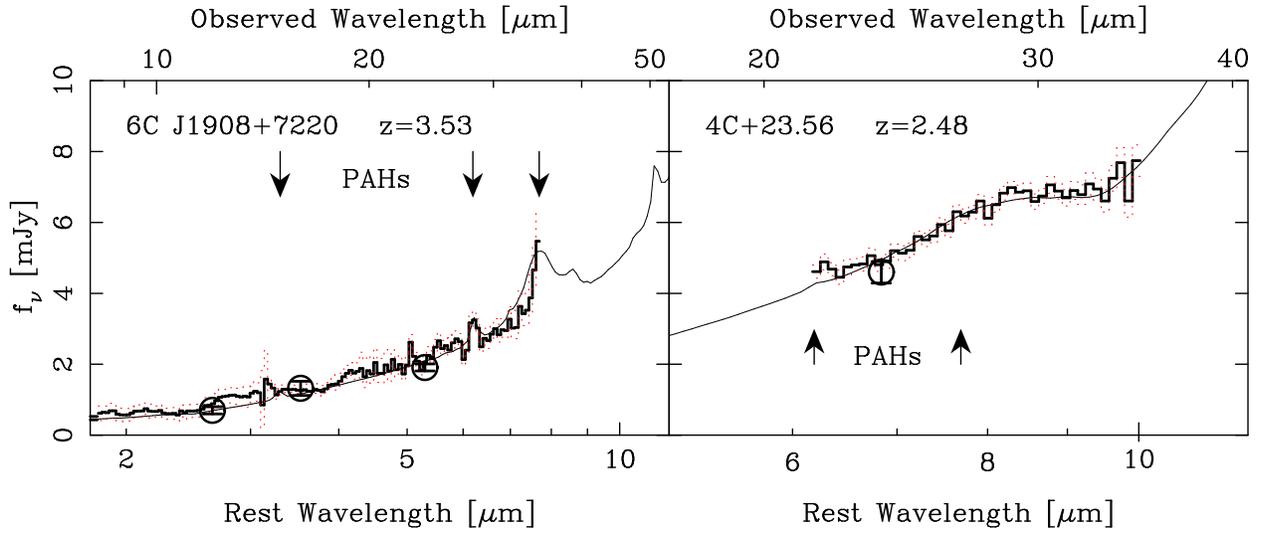}
\caption{IRS spectra of 6C\,J1908+7220 (left) and 4C\,+23.56 (right) marked by thick solid 
  line and $1\,\sigma$ uncertainties indicated by dashed red line. {\it Spitzer} 
  photometry is also plotted along with the model (thin line) fit to the $\sim 1-1000\,\um$ SED 
  shown in Figure~\ref{fig.full}. 
  The vertical arrows indicate the expected location of the strongest PAH 
  features in the wavelength range covered by the IRS spectra (3.3, 6.2 and 
  $7.7\,\um$). Note that 6C\,J1908+7220 exhibits PAH emission, while
  4C\,+23.56 appears to show some $9.7 \mu$m absorption, but exhibits no
  detectable PAH emission.}
\label{fig.irs}
\end{figure*}

\begin{figure*}
\center
\includegraphics[width=12cm,angle=-90]{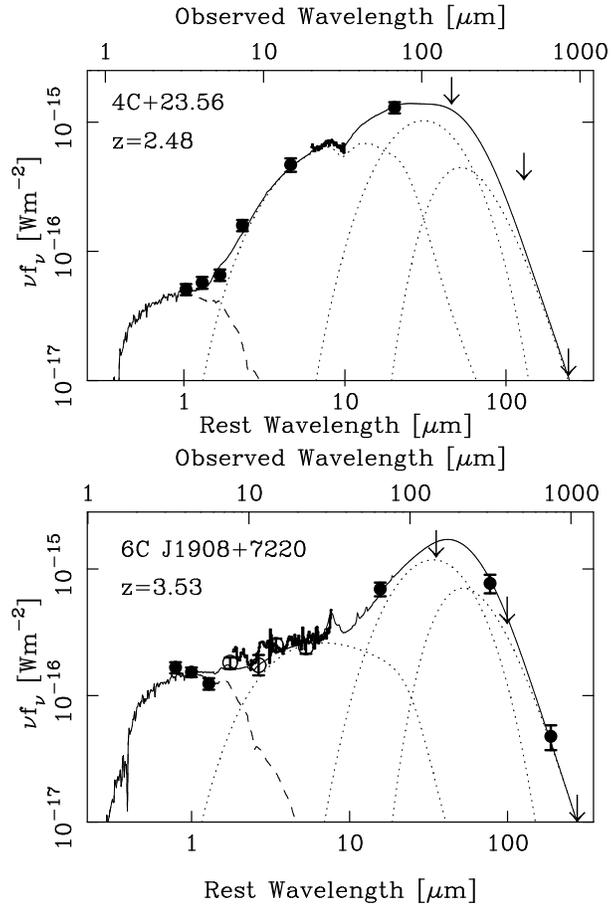}
\caption{Total (1-1000\,$\um$) IR SEDs of 4C\,+23.56 and 6C\,J1908+220 fitted to available 
  data. The fitted data (filled solid points and thick lines) are described in the text. 
  The open points over the same redshift range as the spectra represent data not used in 
  the fitting.
  The model (thin solid line) is comprised of an old stellar component (dashed line) and three 
  dust components (dotted lines) of varying temperatures. In addition, a PAH template 
  is included and the Galactic center extinction curve has been applied.}
\label{fig.full}
\end{figure*}

Figure~\ref{fig.irs} presents the IRS spectra and Figure~\ref{fig.full}
presents full infrared spectral energy distributions (SEDs) of both HzRGs
using all available data from $3.6\,\um$ to $1250\,\um$
(see Table~\ref{tab.data}). Our $16\,\um$ flux densities are updated from 
\citet{Seymour:07} due to newer, improved calibrations for the observing mode
used. The spectra agree well with the photometric data at the same wavelength.

Both 4C\,+23.56 and 6C\,J1908+7220 show strong continuum luminosities consistent 
with the very high, monochromatic rest-frame $5\,\um$ luminosities, 
$L_{5\um}\sim7\times10^{12}L_\odot$ ($\sim 5-10\times10^{46}\,\ergps$), 
presented by \citet{Seymour:07}. However, these two 
HzRGs differ in the strength of their $6.2-7.7\,\um$ PAH complex --- which is clearly 
seen in 6C\,J1908+7720, despite the presence of a known powerful AGN, 
but is undetected in 4C\,+23.56. 
PAH emission in a powerful radio galaxy is unusual; lower
redshift radio galaxies and other AGN generally have much weaker PAH emission 
\citep[e.g.][]{Haas:05,Ogle:06,Cleary:07, Shi:07}. In fact, the integrated PAH
luminosity of 6C\,J1908+7720 is the highest of any source we are aware of,
though the strong mid-infrared continuum conspires to make the observed
$6.2\,\um$ PAH feature equivalent width rather modest.

The mid-infrared spectrum of 4C\,+23.56 is quite different.  Rather than 
showing PAH emission features, the observations of 4C\,+23.56 extend
longward enough to show a flattening at rest-frame wavelengths greater 
than $\sim8.3\,\um$ which we interpret as the onset of the $9.7\,\um$ 
silicate absorption feature. The steep slope of the continuum from 24 to $70\,\um$, 
$S_\nu\sim\nu^{-2}$, for both sources is confirmed by the high measured flux 
densities at $70\,\um$. 
The full IR SED fitting, described 
in the following section, finds a silicate absorption of 
$\tau_{\rm 9.7\um}$=$0.3\pm0.05$ for 4C\,+23.56. This value is high compared 
to most lower redshift radio galaxies of comparable luminosity 
\citep{Ogle:06,Cleary:07}, but low compared to the sources in 
\citet{Sajina:07b} which have far lower radio luminosities, but reside at 
similar redshifts. The redshift 
of 6C\,J1908+7220 makes it impossible to derive a measure of the silicate 
absorption feature from {\it Spitzer}/IRS spectroscopy.

\section{Full Infrared SED fitting}

Here we
supplement our {\it Spitzer} observations which cover the $3.6-160\,\um$ range
with observations from SCUBA, CSO and IRAM which provide detections and upper 
limits for the $350-1250\,\um$ range.
These data are presented in Table~\ref{tab.data} and the references are given therein.
We use the same model 
to fit these data as that used by \citet{Sajina:07a} and \citet{Lacy:07} 
which includes five components: an old (5\,Gyr) stellar component, three dust 
components of different temperatures (hot, cold and warm) and a PAH model. 
The age of the stellar population is unimportant as the wavelength range used in this 
fitting (the near-IR) varies very little for populations older than 1\,Gyr. 
The far-infrared (far-IR) emission 
was fit as a modified cold blackbody with temperature $50$\,K. A warm (small grain) 
component was included with a power-law index of $2$ and cutoffs at high and low 
frequency. The far-IR starburst luminosity, $L_{\rm FIR}$, was calculated 
simply from the cold dust component. 
The hot dust component, associated with Very Small Grains, was 
based on a power-law with variable spectral index and was 
exponentially cut off at high frequencies with a variable cutoff frequency. 
The AGN luminosity, $L_{\rm AGN}$, was calculated from summing the hot and warm components. 
The hot dust component is then reddened by a dust screen assumed to have the Galactic 
center extinction law of \citet{Chiar:06}, likely to be a good approximation to the 
extreme density environments of AGNs \citep{Sajina:07a}. Geometries other than a simple 
screen may be
less efficient at obscuring, hence this $L_{\rm FIR}$ estimate is a likely lower limit. 
We then use a PAH model template based on 
observations of NGC 7714 which we find to have no significant difference to 
M82 \citep{Sajina:07a}.
During the fitting several components were fixed: the amplitude of the stellar 
component was taken from Seymour et al. (2007), and the short-wavelength 
cut-off of the hot component
and the slope of the warm component were also fixed.
For 4C\,+23.56 the cold dust component
was fixed to the maximum allowed value and for 6C\,J1908+7220 the reddening was
fixed to zero as we could not constrain it via the silicate absorption
feature.

The principle difference in the observed SEDs, presented in Figure~\ref{fig.full}, 
is that 6C\,J1908+7220, with it's strong PAH
emission features, is detected at both $350\,\um$ from CSO/SHARCII and at 
$850\,\um$ from JCMT/SCUBA in contrast to 4C\,+23.56 (which has not been observed at 
$350\,\um$ or detected in SCUBA observations). While the non-detection of 4C\,+23.45 with 
SCUBA may be due to its lower redshift (and the negative k-correction at sub-mm 
wavelengths), the total far-IR luminosity is constrained to 
be significantly less than that of 6C\,J1908+7220. The results of our fitting for both 
HzRGs are presented in Table~\ref{table.results},  including total far-IR luminosities, 
inferred AGN luminosities, PAH strengths, dust mass, optical depth, and inferred SFR. 
For 4C\,+23.56 many of these quantities are  upper limits. 

In addition to the detections at $24$ and $850\,\um$, the upper limits at 
$450$ and $1000\,\um$ provide strong constraints on the 
amount and temperature of cold dust in 6C\,J1908+7220. We obtain a starburst far-IR 
luminosity of $2.3\pm0.4\times10^{13}\,L_\odot$ ($\sim4000\pm700M_\odot$yr$^{-1}$, Kennicutt 1998) 
where the modified black-body cold dust parameters 
are constrained to be $T_c$=$50\pm 5\,$K and $\beta$=$1.5 \pm 0.2$ following the approach 
of \citet{McMahon:99}. This result implies
$\sim8\pm2\times 10^8\,M_\odot$ of cold dust. We find an approximate upper-limit 
to the cold dust in 4C\,+23.56 of $\la1.6\times10^8\,M_\odot$ for $T_c$=$50\,$K.
Using the \citet{Pope:07} relations, the SFR in 6C\,J1908+7220
inferred from our $6.2\,\um$ PAH luminosity
is $\sim8000\,M_\odot$yr$^{-1}$, around a factor of two higher than that inferred from 
the far-IR luminosity, albeit with considerable uncertainty
due to the scatter in this relation.
We are able to provide upper limits to the $6.2\,\um$ PAH and the far-IR (starburst) 
luminosity for 4C\,+23.56.
These limits are relatively high, hence the SFR in 4C\,+23.56
could still be at a level considered extreme in the local Universe, 
$\sim1000\,M_\odot$yr$^{-1}$. 

\section{Interpretation and Conclusions}

\begin{figure}
\center
\includegraphics[width=8cm,angle=0]{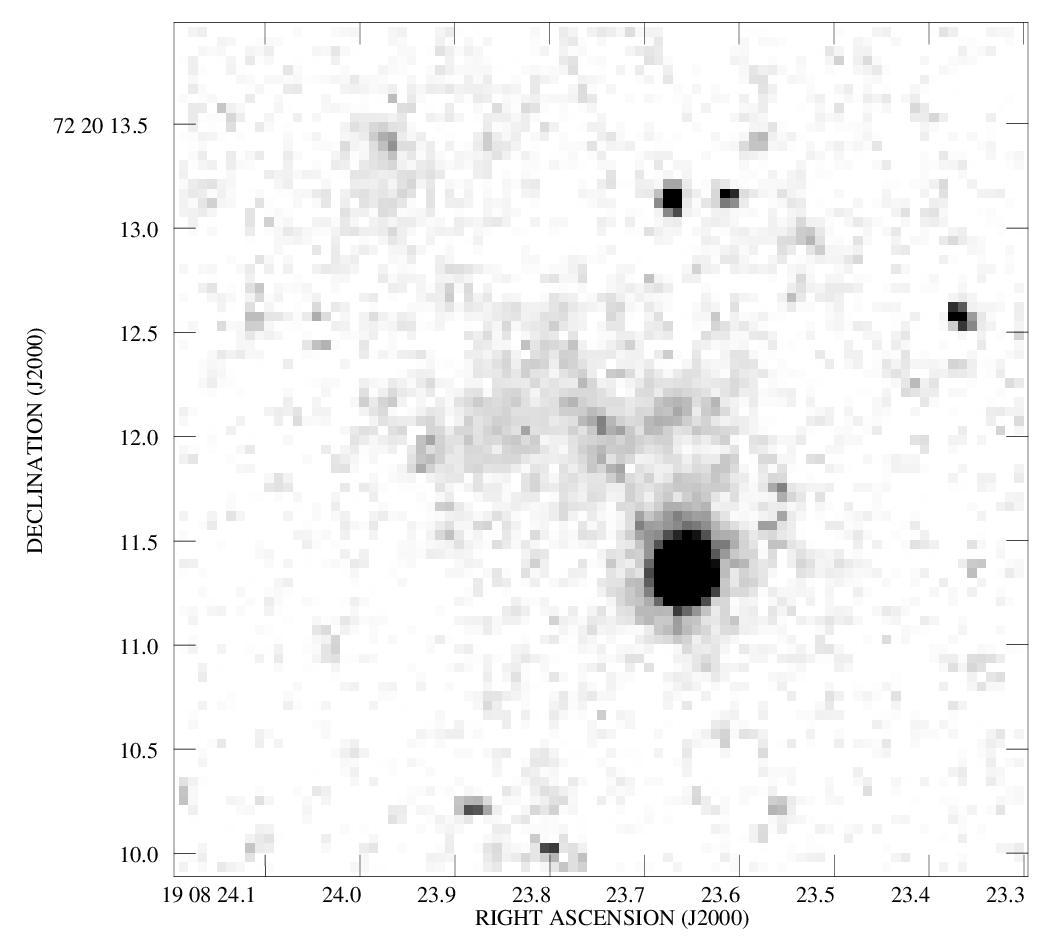}
\caption{Image of 6C\,J1908+7220 obtained with {\it HST}/WFPC2 in the F702W band. 
  Note the extended, diffuse emission to the NE
  at these rest-frame far-UV wavelengths, likely associated with extended
  star formation.  
  The IRS observations with a wide slit completely cover the extended emission.}
\label{fig.hst}
\end{figure}

The clear presence of PAHs in 6C\,J1908+7220, a known powerful AGN, is consistent 
with results at lower redshift where PAH emission is seen in AGN and quasar 
host spectra \citep{Shi:07}. PAHs can survive the high radiation environment of the AGN if
shielded by an obscuring torus or located far enough from the AGN that 
they are simply stimulated and not destroyed. An {\it HST} WFPC2 image of 
6C\,J1908+7220  taken with the F702W filter (PI G. Miley) indicates extended, 
irregular star formation in the rest-frame far UV supporting the view of star formation 
occurring away from the nucleus in an irregular morphology (see Figure~\ref{fig.hst}). 
The absence of PAH emission from 4C\,+23.56 is potentially due to destruction by the 
AGN, but could also simply be due to the relatively lower intrinsic SFR. No 
{\it HST}\, imaging of sufficient quality is available for this source, but it does 
possess a relatively large Ly$\alpha$ halo \citep{VillarMartin:07} implying at least a 
reasonably large reservoir of material for star formation.

Both HzRGs have deep UV spectropolarimetry
\citep{Dey:99c, Vernet:01}. 4C\,+23.56 is highly polarized ($P$=$15.3\pm2\%$), 
while 6C\,J1908+7220 is unpolarized ($P$$<$$5\%$) and shows broad absorption lines. 
The contribution of a scattered AGN is therefore expected to be much higher in 
4C\,+23.56 than in 6C\,J1908+7220. As 4C\,+23.56 is almost two magnitudes fainter 
than 6C\,J1908+7220 at optical wavelengths
\citep[$R$=$23.2$ for 4C\,+23.56 compared to $R$=$21.4$ for 6C\,J1908+7220 $-$][]{Chambers:96, Pentericci:00},
the remaining rest-frame continuum emission (i.e. UV radiation from hot stars), 
after correction for the scattered 
AGN emission, is at least an order of magnitude fainter in 4C\,+23.56 than in 
6C\,J1908+7220. While the determination of SFRs from the rest-frame 
UV continuum remains very uncertain in these objects, this large difference 
does suggest a much higher SFR in 6C\,J1908+7220,
consistent with our IRS results.

The mid-IR luminosities and 70/$24\,\um$ flux density ratios (roughly equivalent 
to 15/$7\,\um$ rest-frame) of these two HzRGs are at the high end of the 
distribution of the radio galaxies studied by \citet{Ogle:06} and \citet{Seymour:07}.
However, there is a marked difference in their 8 to $24\,\um$ slope, 
beyond the fact they sample slightly different rest frame wavelengths.
The steeper 8 to $24\,\um$ slope of 4C\,+23.56 indicates a much more
powerful AGN relative to the host galaxy rest-frame near-IR luminosity (and hence 
stellar mas) compared to 6C\,J1908+7220.

The mid-IR luminosities and slopes of these sources are also similar to high 
redshift $24\,\um$ selected ULIRGs studied by several groups \citep[\eg][]{Yan:07, 
Dey:08}. This similarity is not surprising given the selection of those samples 
(very red from $R-$band to $24\,\um$ and $S_{\rm 24um}\ga 1$\,mJy), but does 
support our earlier comment that HzRG are representative of a larger population 
which experience brief periods of extreme AGN activity and obscured star formation. 
In fact, some of these $24\,\um$-selected ULIRGs do host radio-loud AGN 
\citep{Sajina:07b}, but have radio luminosities several orders of magnitudes less 
than the sources presented here. Additionally, the correlation between the presence 
of PAH and strong emission from cold dust has been seen in {\it Spitzer} selected 
type 2 AGN (Lacy et al. 2007; Sajina et al. 2008). 
Further IRS and far-IR observations would provide us with an excellent 
opportunity to examine how star formation
is related to the obscuration of powerful AGN and their radio-loud phase.

\acknowledgments

We thank the referee for help improving the presentation of this paper.
This work is based
on observations made with the {\it Spitzer Space Telescope}, which
is operated by the Jet Propulsion Laboratory, California Institute of
Technology under a contract with NASA. Support for this work was provided
by NASA through an award issued by JPL/Caltech.

{\it Facilities:} \facility{{\it Spitzer} (IRS)}.

\bibliographystyle{apj.bst}


\begin{deluxetable}{cccc}
\tablecaption{Photometry used in the SED fitting.}
\tablewidth{0pt}
\tablehead{
\colhead{HzRG} &
\colhead{wavelength} &
\colhead{flux density} & 
\colhead{reference}}
\startdata
                & $\um$ & $\uJy$  &   \\
\hline\
6C\,J1908+7220  & 3.6 & $200\pm20$ & 1\\
                & 4.5 & $229\pm23$& 1 \\
               & 5.8 & $241\pm25$& 1\\
               & 8.0 & $480\pm48$ & 1\\
               & 12  & $840\pm100$  & 2\\
               & 16  & $1320\pm70$ & 3\\
               & 24  & $1910\pm100$  & 1\\
               & 70  & $16200\pm1900$  & 1\\
               & 160 & $<63300$       & 1\\
               & 350 & $90000\pm15000$     & 4\\
               & 450 & $51000$       & 5\\
               & 850 & $13500$    & 6\\
               & 1250 & $3000$         & 6\\
               &        &            &  \\
4C\,+23.56      & 3.6  & $61\pm6$& 1\\ 
                & 4.5  & $86\pm9$& 1   \\
                & 5.8  & $127\pm13$&1\\
                & 8.0  & $424\pm043$&1\\
                & 16   & $1610\pm130$ &2\\
                & 24   & $4390\pm200$ &1\\
                & 70   & $30300\pm3000$  &1\\
                & 160  & $<70500$ &1\\
                & 450  & $<51000$ &7\\
                & 850  & $<2940$ &7\\
\enddata
\label{tab.data}
\tablerefs{(1) \citet{Seymour:07}; (2) \citet{Siebenmorgen:04}; (3) this work;
  (4) \citet{Greve:06a}; (5) \citet{Reuland:04}; (6) \citet{Papadopoulos:00};
  (7) \citet{Archibald:01}}
\end{deluxetable}

\clearpage
\thispagestyle{empty}
\setlength{\voffset}{15mm}
\begin{deluxetable}{cccccccccc}
\small
\tablecaption{Measured Global Properties of the HzRGs From the IR SED Fits}
\tablewidth{0pt}
\tablehead{
\colhead{HzRG} &
\colhead{z} &
\colhead{$L_{\rm 500MHz}$} &
\colhead{$L_{\rm FIR}$} &
\colhead{$L_{\rm AGN}$} &
\colhead{$L_{\rm 6.2\um}$} &
\colhead{EW$_{6.2\um}$ } &
\colhead{$\tau_{\rm 9.7\,\um}$} &
\colhead{$M_{\rm dust}$} &
\colhead{SFR$_{FIR}$}}
\startdata
               &     & ergs$^{-1}$Hz$^{-1}$ & $L_\odot$ & $L_\odot$ & $L_\odot$ & $\um$ &    & $M_\odot$       & $M_\odot$yr$^{-1}$\\
\hline\
6C\,J1908+7220 & 3.53 & $7.9\times10^{28}$ & $2.3\pm0.4\times10^{13}$& $6,3\pm1.9\times10^{13}$& $8.3\pm2.0\times10^{10}$& $0.04\pm0.01$ & n/a\tablenotemark{a}& $8\pm2\times10^8$ & $4000\pm700$\\
4C\,+23.56     & 2.48 & $1.3\times10^{29}$ & $\la6.2\times10^{13}$&$\la3.8\times10^{13}$& $\la1.1\times10^{10}$ & n/a\tablenotemark{b} & $0.3\pm0.05$&$\la1.6\times10^8$ & $\la100$\\
\enddata
\label{table.results}
\tablenotetext{a}{The $9.7\,\um$ silicate feature is not covered by current observations.}
\tablenotetext{b}{The $6.2\,\um$ PAH feature is not detected in this source.}
\end{deluxetable}
\setlength{\voffset}{0mm}

\clearpage


\end{document}